\documentclass[12pt]{iopart}
\usepackage{epsf}

\begin{document}
\title{One dimensional lattice of permanent magnetic microtraps for ultracold atoms on an atom chip}
\author{Mandip Singh, Michael Volk, Alexander Akulshin, Andrei Sidorov, Russell McLean and Peter Hannaford}

\address{Centre for Atom Optics and Ultrafast Spectroscopy, ARC Centre of Excellence for Quantum Atom Optics,
 Swinburne University of Technology, Melbourne, 3122, Australia }

\ead{masingh@swin.edu.au}
\begin{abstract}
  We report on the loading and trapping of ultracold atoms in a one dimensional permanent magnetic lattice of period 10 $\mu$m produced on an atom chip. The grooved structure which generates the magnetic lattice potential is fabricated on a silicon substrate and coated with a perpendicularly magnetized multilayered TbGdFeCo/Cr film of effective thickness 960 nm. Ultracold atoms are evaporatively cooled in a Z-wire magnetic trap and then adiabatically transferred to the magnetic lattice potential by applying an appropriate bias field. Under our experimental conditions trap frequencies of up to 90 kHz in the magnetic lattice are measured and the atoms are trapped at a distance of less than 5 $ \mu$m from the surface with a measured lifetime of about 450 ms. These results are important in the context of studies of quantum coherence of neutral atoms in periodic magnetic potentials on an atom chip.

\end{abstract}

\section{Introduction}

 Optical lattices based on the interference of intersecting laser beams \cite{Blo05} provide controllable periodic potentials for the manipulation of ultracold atoms. The tunnelling amplitude between adjacent potential wells can be adjusted by varying the barrier height allowing one to establish analogies with condensed matter physics and to explore many interesting phenomena. Spin dependent coherent transport of ultracold atoms \cite{Olaf032} and the realization of multiparticle entanglement \cite{Olaf03} are among the phenomena with implications for quantum information processing \cite{Treut06} that have been studied in optical lattices. As well, the study of quantum phase transitions such as the superfluid-to-Mott-insulator transition can be conveniently explored in optical lattices \cite{Gre02}.

A periodic array of current carrying wires or permanent magnets creates an exponentially decreasing magnetic field from the surface \cite{Opat92} which has been used to reflect cloud of cold atoms from magnetic mirrors \cite{Roach95, Sido96}. An additional uniform magnetic field perpendicular to the array produces a periodic pattern of magnetic field minima \cite{Opat92} which can be regarded as a one dimensional magnetic lattice \cite{Hinds99,Ghan06} where atoms in weak field-seeking states can be trapped at the magnetic field minima. In contrast to an optical lattice, a magnetic lattice does not require any laser beams or associated stabilization and there is no spontaneous emission involved in the trapping process. Since the required periodic field can be readily generated by means of a magnetic microstructure, a magnetic lattice can be easily integrated on an atom chip \cite{Fol02, For07}. The period of the magnetic lattice depends on the fabrication parameters of the microstructure, whereas in the case of an optical lattice it is normally half of the laser wavelength. The flexibility of being able to choose larger periods in the case of a magnetic lattice provides an easy way to resolve atoms in individual lattice sites. A magnetic lattice can be generated using either current carrying wires or permanent magnetic materials.  A lattice generated from microstructured
permanent magnetic materials is expected to have higher intrinsic stability in the context of trap lifetime and lower heating rates, due to suppression of technical noise, compared with current carrying wire microtraps under similar conditions. Since atoms in weak field-seeking states can be trapped magnetically and trap parameters such as trap frequency are spin state dependent, this property provides a way of state dependent manipulation of ultracold atoms in a magnetic lattice. Furthermore, the magnetic lattice potential can be characterized with rf spectroscopy techniques \cite{Whit07} and also there is the potential advantage of being able to use rf evaporative cooling techniques in a magnetic lattice. In the context of quantum information processing the realization  of a  quantum gate demands a coherence time longer than the gate operation time. Coherence times of the order of 1 s have been observed on an atom chip with atoms at a distance of about 5  $\mu$m from the surface for two different hyperfine states of $^{87}$Rb \cite{Reich02, Treut04}. The effect of the surface on the lifetime of atoms in a magnetic microtrap is discussed in \cite{Fort02}. Finally, a magnetic lattice may be regarded as a promising candidate for the possible realization of a scalable quantum computer \cite{Cirac00, Rauss01}.

In this paper we report the trapping of ultracold  $^{87}$Rb atoms in a one-dimensional permanent magnetic lattice of period 10 $\mu$m on an atom chip. We describe the design and construction of the magnetic lattice atom chip and the loading of ultracold atoms into the magnetic lattice. Our atom chip is fabricated with a hybrid technology \cite{Hall06} and consists of current carrying wires and a perpendicularly magnetised grooved structure. The current carrying wires are used for trapping and cooling of ultracold atoms with standard atom chip techniques and for loading the magnetic lattice and a micro-structure to generate the magnetic lattice potential is constructed on a silicon substrate and coated with a multilayered magnetic film. Preliminary results of this work have been reported previously \cite{Sin07}. A two-dimensional FePt magnetic lattice with period of 20 $\mu$m has recently been reported and trapping of ultracold $^{87}$Rb atoms was demonstrated in $>$ 30 lattice sites \cite{Gerr07}.

\section{One Dimensional Periodic Array of Magnetic Potentials}

\begin{center}
\begin{figure}
\begin{center}
\epsfysize=60mm
\epsfbox{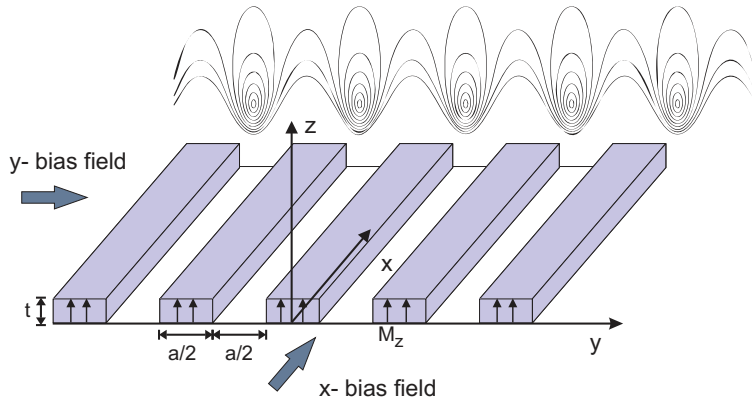}
\end{center}
\caption{\label{fig:1} Schematic showing an array of perpendicularly magnetized parallel slabs with a period of 10 $\mu$m. A magnetic lattice of elongated microtraps with non-zero potential minima is formed by applying a bias field along the $y$-direction and an additional bias field along the $x$-direction.}
\end{figure}
\end{center}

      In our experiment a periodic array of magnetic potential minima is produced from a perpendicularly magnetized pattern of parallel uniformly spaced slabs of period $a$, spacing $a/2$, and thickness $t$ by applying appropriate bias fields as shown in Figure~\ref{fig:1}. For distances from the surface large compared to $ a/4\pi$ and for perpendicular magnetization $M_{z}$ the field components can be written as

      \begin{equation}
      \label{eq:Bc1}
      B_{x}= B_{bx}
      \end{equation}

       \begin{equation}
       \label{eq:Bc2}
      B_{y} = B_{0}\sin(ky)e^{-kz} + B_{by}
      \end{equation}

       \begin{equation}
      \label{eq:Bc3}
      B_{z} = B_{0}\cos(ky)e^{-kz} + B_{bz}
        \end{equation}

    where $k = 2\pi/{a}$,  $ B_{0}= B_{s}(e^{kt}-1)$,  $B_{s}= 4M_{z}$ (Gaussian units) and $B_{bi}$ for $i= x,y,z$ are the bias field components. In the absence of bias fields the
    magnitude of the magnetic field is given by

      \begin{equation}
      \label{eq:Bc4}
      B(x,y)= B_{0}e^{-kz}
      \end{equation}

       i.e, the magnitude of the field decreases exponentially with distance $z$ from the surface. The field gradient repels atoms in weak field-seeking states and such a structure behaves as a magnetic mirror. By applying a bias field in the $y$- or $z$-direction one can induce corrugation in the field magnitude which can give rise to a one-dimensional array of magnetic field minima with nonzero potential, given by $|B_{bx}|$. For $B_{bz} =0$ and bias fields along the $x-$ and $y$-direction, the loci of the field minima (trap axis) are given by
        \begin{equation}
         \label{eq:Bc5}
      y_{min}=\left(n_{y}-\frac{1}{4}\right)a
      \end{equation}
          \begin{equation}
    \label{eq:Bc6}
      z_{min}= \frac{a}{2\pi}\ln\left(\frac{B_{0}}{|B_{by}|}\right)
      \end{equation}
     where $n_{y}=0, \pm 1, \pm 2,....$ and each value of $n_{y}$ corresponds to a different trap centre along the $y$-direction.

      Now according to the symmetry the field along the trap axis (trap bottom) is the same as  $ |B_{bx}|$ and the trap depth along the confining directions is given by \cite{Ghan06}

      \begin{equation}
    \label{eq:Bc7}
     \Delta B_{y} = (B^{2}_{bx} + 4B^{2}_{by})^{1/2}-|B_{bx}|
           \end{equation}
and
    \begin{equation}
    \label{eq:Bc8}
     \Delta B_{z} = (B^{2}_{bx} + B^{2}_{by})^{1/2}-|B_{bx}|
  \end{equation}

  The trap bottom can be chosen appropriately by adjusting $ B_{bx}$ independently in order to prevent loss due to Majorana spin flips. The curvature of the field magnitude at the trap axis along the confining directions determines the trap frequencies which for an atom of mass $m$ and Land\'{e} g factor $g_{F}$ in a magnetically trappable state with magnetic quantum number $m_{F}$ is given by \cite{Ghan06}
\begin{equation}
    \label{eq:Bc9}
     \omega_{y}=\omega_{z}=\frac{2\pi}{a}\left(\frac{m_{F} g_{F} \mu_{B}}{m |B_{bx}|}\right)^{1/2} |B_{by}|
  \end{equation}
  where $\mu_{B}$ is the Bohr magneton.

  For a fixed lattice period and a given magnetic quantum number $m_{F}$ the trap frequencies, unlike the trap position, are a linear function of the bias field $B_{by}$. The dependence of the trap depth (barrier height) on the bias fields should allow control of the tunnelling amplitude between adjacent sites. However, for larger periods one cannot asymptotically lower the barrier height and decrease the trap frequencies because at some point the magnetic field gradient will not be sufficient to hold the atoms against gravity and eventually the atoms will drop out of the trap if the lattice is mounted up-side down, which is normally the case in order to perform time of flight measurements.

\section{The Magnetic Lattice}

\begin{center}
\begin{figure}
\begin{center}
\epsfysize=80mm
\epsfbox{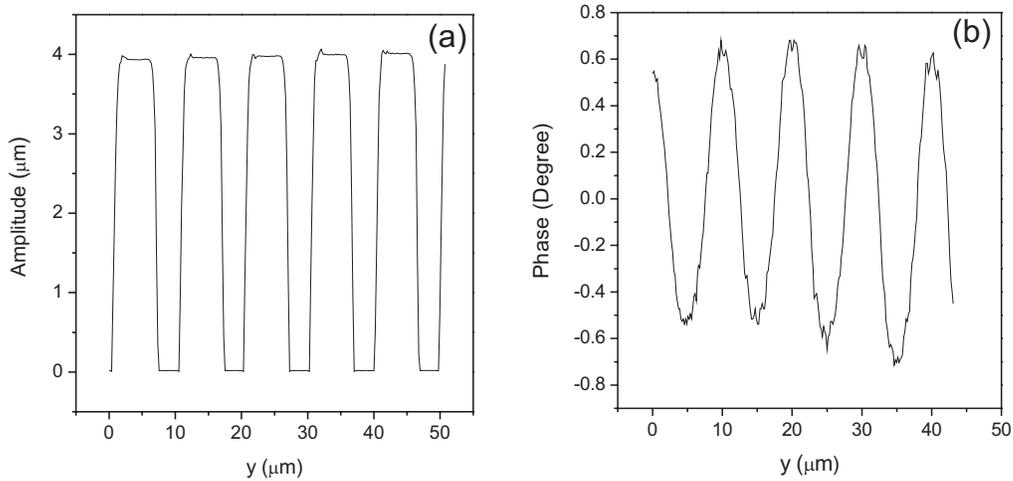}
\end{center}
\caption{\label{fig:2}(a) Atomic force microscope measurement of the grooved structure, and (b) Magnetic force microscope measurement indicating the coated micro-structure is magnetized and the field undulation has the same period as that of the grooves.}
\end{figure}
\end{center}

   The micro-structure required to generate a magnetic lattice potential was fabricated on a silicon wafer (35 mm $\times$ 35 mm) of thickness 300  $\mu$m. The wafer was selectively etched by reactive ion etching in order to fabricate a grooved structure of period 10 $\mu$m and depth $>$ 50 $\mu$m where the length of each groove is 10 mm and a total of one thousand parallel grooves are etched.  A multilayered magneto optical Tb$_{6}$Gd$_{10}$Fe$_{80}$Co$_{4}$ film  \cite{Wang05} was coated on the grooved structure with a thin film deposition system (Kurt J Lesker CMS-18) at a temperature of $100^{0}$C to produce the required magnetic pattern to generate a magnetic lattice. Composite TbGdFeCo and chromium targets were used to coat the film by DC magnetron sputtering. The magnetic anisotropy of the film was found to deteriorate if its thickness was greater than 250 nm. To circumvent this problem we used a multilayer structure. First a 100 nm thick chromium film was deposited on the silicon grooved structure followed by a 160 nm thick TbGdFeCo magneto optical film. A total of six such layers were coated to give an effective thickness of 960 nm of magneto-optical film. A 10 nm chromium film was then coated on top of  this multilayer structure and finally a gold film of 150 nm was deposited in order to have good optical reflectivity for the mirror MOT. The anticipated Curie temperature of the Tb$_{6}$Gd$_{10}$Fe$_{80}$Co$_{4}$ film is about $300^{0}$C; a high Curie temperature is required to prevent demagnetization during bake-out. Other parameters which determine the film quality include coercivity ($H_{c}$), remanent magnetization ($M_{z}$) and  saturation magnetization ($M_{S}$). High coercivity is required to prevent loss of magnetization under the action of large bias and other magnetic fields. All of these qualities are acquired by Tb$_{6}$Gd$_{10}$Fe$_{80}$Co$_{4}$  magneto optical films. The coated silicon structure was analysed under an atomic force microscope (AFM). Figure~\ref{fig:2}(a) shows the grooved structure with the 10 $\mu$m period of the structure and a measured rms roughness of 20 nm. The magnetic force microscope (MFM) profile in Figure~\ref{fig:2}(b) indicates that the structure is periodically magnetized with a period the same as that of the grooves. The magnetic properties, including coercivity and remanent magnetization of a multilayer magneto-optical film prepared under similar conditions on a silicon substrate were analysed by SQUID measurement. The SQUID data shows that the coercivity of our film is about 6 kOe and the remanent magnetization ($4 \pi M_{z}$) about 3 kG. This data also indicates that the remanent magnetization of the sample is close to the saturation magnetization ($M_{S}$), which reflects the good magnetic homogeneity of the film.

\section{Atom Chip Construction}

\begin{center}
\begin{figure}
\begin{center}
\epsfysize=100mm
\epsfbox{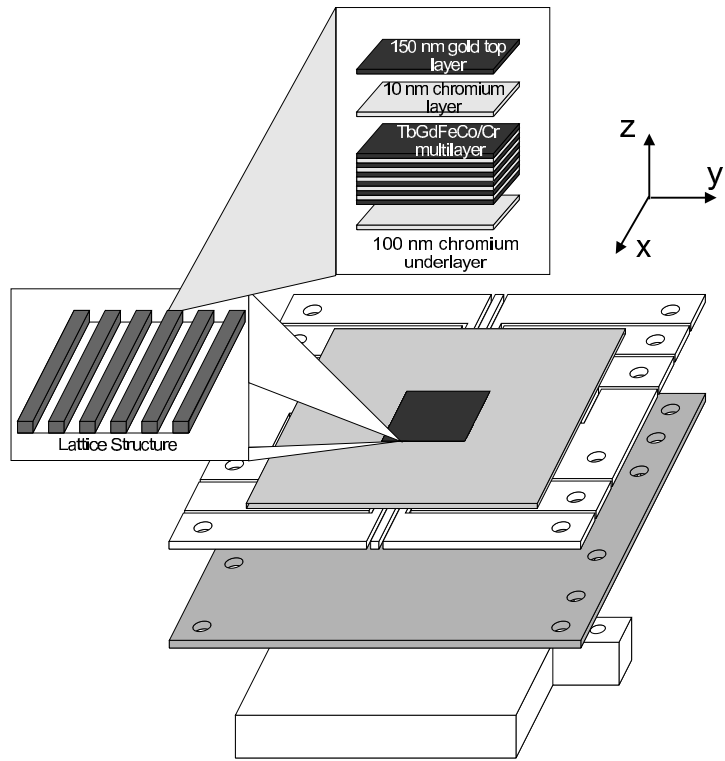}
\end{center}
\caption{\label{fig:3} Construction of the periodic magnetic micro-structure on the atom chip.}
\end{figure}
\end{center}

The atom chip consists of a silver foil which was glued on a machinable ceramic (50 mm $\times$ 50 mm)  with a high vacuum thermal conductive epoxy (EPO-TEK H77). The atom chip construction is shown in Figure~\ref{fig:3}. The epoxy was heated to $150^{0}$C for one hour so that it turns into a glassy state to form a strong bond. A thin chromium film was also deposited on one side of the silver foil in order to increase the bond strength with the epoxy. The silver foil was then micro-machined  to obtain the required pattern of the current carrying wires. The length of the Z-wire was about 5 mm with thickness 0.5 mm and width 1 mm. The patterned silver foil was then polished, cleaned and heated to 100$^{0}$C in order to remove trapped water and other cleaning agent vapours.

  The silicon wafer of thickness 300 $\mu$m  containing the periodic structure coated with the multilayer magnetic thin film was then glued with the same high vacuum epoxy and heated carefully to avoid temperature gradients across the silicon wafer. The whole chip assembly was then baked at a maximum temperature of 95$^{0}$C to avoid any temperature induced degradation of the magnetic film.

\section{Loading Atoms in the Lattice Potential}

 Using standard laser cooling and trapping techniques on an atom chip \cite{Hall06} we transfer about $1 \times10^{8}$  $^{87}$Rb atoms from a mirror magneto-optical trap (MOT) to a compressed MOT generated by a current (22 A) flowing through a U-shaped wire and a bias field of about 7 G. Diffraction of the trapping light from the grooved structure reduces the reflection efficiency to only 40\% from the grooved area and limits the quality of the mirror MOT. The atoms are further cooled to 40 $\mu$K through polarisation gradient cooling by red detuning the laser by 55 MHz for 40 ms. The U-wire current is then switched off and an optical pumping pulse is applied for 1.5 ms to transfer atoms to the $|F = 2, m_{F} = +2\rangle$ low field-seeking state. After this process the Z-wire trap is turned on and the wire current is ramped from 25 A to 35 A in 200 ms and the bias field $B_{bx}$ used to trap atoms in the Z-wire trap is simultaneously ramped from 15 G to 40 G. In this way we can trap $6\times10^{7}$ atoms at a temperature of about 200 $\mu$K with a trap lifetime of about 40 s. Radiofrequency evaporative cooling is started after waiting 500 ms for thermalization of the trapped atoms. The trap bottom is adjusted by applying a bias field along the $y$-direction.
 We are able to create a BEC in the Z-wire trap (at a distance from the surface of about 200 $\mu$m where the lattice effect is negligible) by applying a logarithmic sweep from 20 MHz to 730 kHz for 10 s keeping the Z-wire trap parameters fixed. To load atoms into the magnetic lattice, however, we start with an ultracold atom cloud rather than a BEC. The final radiofrequency is stopped at 2 MHz in a 10 s sweep to obtain an ultracold cloud of $7\times10^{6}$ atoms at a temperature of 15 $\mu$K. In order to move the cloud closer to the lattice the Z-wire current is ramped down linearly to a final value of 5 A in 80 ms. During this phase the $y$-bias field is switched off to lift the trap bottom to 6.9 MHz to prevent Majorana spin flips. After ramping the Z-wire current down to 5 A the bias field along the $y$-direction is ramped up to 30 G in 30 ms keeping the Z-wire current at 5 A.  The action of applying the $y$-bias field creates a periodic trapping lattice potential and since the cloud overlaps with the lattice potential this process results in the loading of atoms into the lattice potential. After the loading process the Z-wire current is reduced to zero and the bias field $B_{bx}$ which now acts as the trap bottom in the lattice is reduced to 15 G. The field along the $y$-direction can also be adjusted accordingly to change the trap frequency and trap depth. We observed a 45\% loading efficiency from the Z-wire trap into the magnetic lattice.
\begin{center}
\begin{figure}
\begin{center}
\epsfysize=80mm
\epsfbox{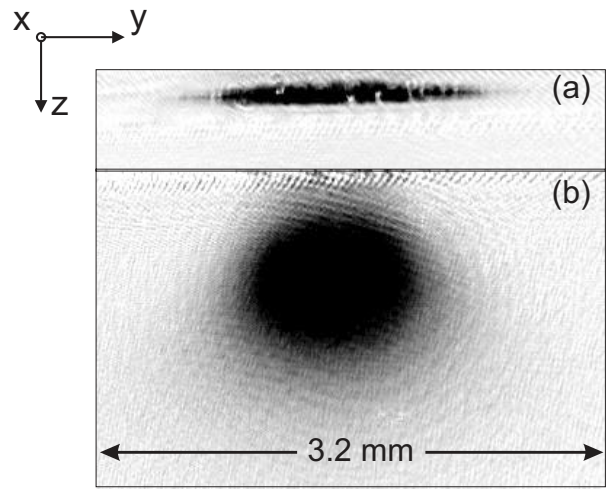}
\end{center}
\caption{\label{fig:4} Absorption image of atoms released from the magnetic lattice after (a) 4 ms and (b) 16 ms time of flight. The broken structure in (a) is due to nonuniform reflection of the imaging beam from the grooved structure.}
\end{figure}
\end{center}

\begin{center}
\begin{figure}
\begin{center}
\epsfysize=80mm
\epsfbox{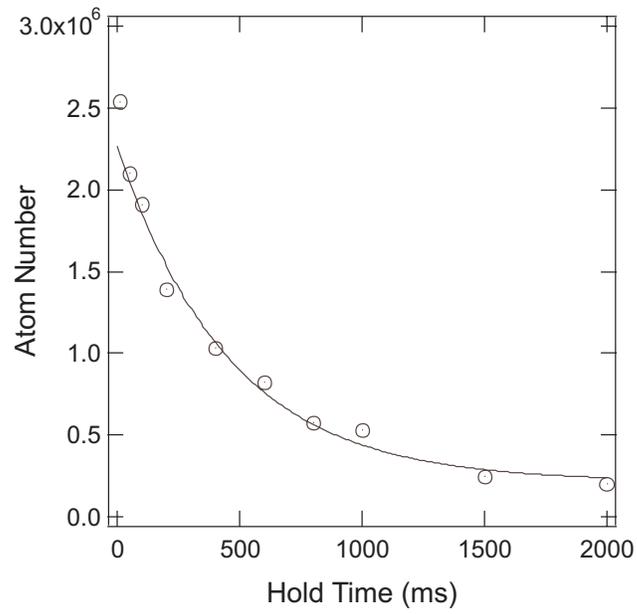}
\end{center}
\caption{\label{fig:5} Atom number remaining in the magnetic lattice as a function of hold time. The solid line represents a fit of an exponential decay of the experimental data points (circles).}
\end{figure}
\end{center}

\begin{center}
\begin{figure}
\begin{center}
\epsfysize=100mm
\epsfbox{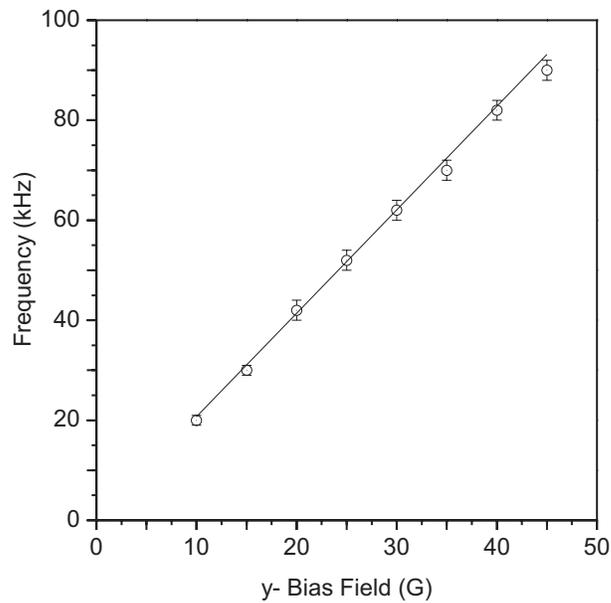}
\end{center}
\caption{\label{fig:6} Radial trap frequency in the magnetic lattice measured via parametric heating. The solid line represents the trap frequency calculated from Eq.~\ref{eq:Bc9}.}
\end{figure}
\end{center}

  We calculated the distance of the trapped atoms from the surface to be 4 $\mu$m under our normal experimental conditions. The cloud of atoms released from the lattice after a 10 ms hold time is shown in Figures~\ref{fig:5}(a) and (b) for 4 ms and 16 ms time of flight, respectively. From the size of the cloud we infer that about 150 lattice sites are occupied after transferring from the 5 mm long Z-wire. We do not resolve atoms in the individual lattice sites because of the low magnification of our detection optics (pixel size = 9 $\mu$m). However, in order to verify that the atoms are actually trapped in the magnetic lattice potential we performed a series of measurements of the lifetime of the trapped cloud and of the trap frequencies. The lifetime of the trapped atoms is measured by holding the atoms for a variable time and looking at the atoms remaining in the lattice potential as shown in  Figure~\ref{fig:6}. The observed lifetime is 452 ms. Because of the symmetry of the magnetic lattice, if the $y$-bias field is reversed then trapping should still be observed. We measured a lifetime of 450 ms when the $y$-bias is applied in the reverse direction. As expected from the symmetry (see Eq.~\ref{eq:Bc2}) trapping was also observed when the trapping bias field was applied along the $z$-direction. Confirmation that atoms are being trapped in the lattice is found by looking at the trap bottom using rf-spectroscopy measurements. We find that our trap bottom is the same as the applied $x$-bias field, consistent with the model in Section 2. No axial confinement was applied to the traps though a weak axial confinement is possible by running current through the Z-wire. However, we estimate a 1 Hz axial trap frequency in the lattice due to confinement caused by the uniform magnetic film on the flat silicon surface surrounding the lattice structure. We performed a measurement of the radial trap frequency in the lattice by parametric heating for a 15 G trap bottom ($x$-bias field) by perturbing the trap position and the trap frequency by applying an additional high amplitude ac-signal on the U-wire for 200 ms. We observed a loss of atoms near the parametric resonance because the magnetic lattice traps are not confined along the axial direction.

   The data are shown in Figure~\ref{fig:6} where the solid line represents the trap frequency calculated from Eq.~\ref{eq:Bc9}. The experimental data show good agreement with the calculated frequency. From the rf-spectrum of the trapped atoms we estimate that the temperature of the trapped cloud is more than 150 $\mu$K as a result of adiabatic compression during transfer from the compressed Z-wire trap to the tightly confining lattice traps. A measurement of the temperature based on time of flight after releasing the trapped atoms from the lattice indicates a temperature of 6 - 8 $\mu$K. The difference in measured values of the temperature in the lattice and after releasing from the lattice is due to adiabatic decompression. We anticipate that it will be possible to lower the temperature of the trapped atoms substantially by increasing the confinement along the axial direction and using rf evaporative cooling.

\section{Summary}

   We have reported the loading of an ultracold cloud of $^{87}$Rb atoms in a magnetic lattice created by a multilayered TbGdFeCo magnetic film coated on a grooved silicon structure of period 10 $\mu$m. We have presented experimental evidence for trapping of atoms in the lattice potential through measurement of the lifetime, trap bottom and trap frequency. Since the period of the lattice is 10 $\mu$m the micro-traps in our experiment are located at a distance of less than 5 $\mu$m from the surface. Under these experimental conditions we measured a lifetime of the trapped atoms in the lattice of about 450 ms. Finally we characterized the magnetic lattice by using parametric heating to determine radial trap frequencies up to 90 kHz, in good agreement with the calculated values. This experiment serves as a platform to realize quantum coherent dynamics in periodic potentials on an atom chip.

 \ack{We would like to thank Brenton Hall for fruitful discussions and James Wang for coating the magneto optical films. We also thank David Gough and Timothy Roach for their contributions in the early stages of this experiment. This work is supported by the Australian Research Council and a Swinburne University strategic initiative grant.


\end{document}